\documentclass[review]{elsarticle}
\usepackage{graphicx}
\usepackage{algorithmic}
\usepackage[]{algorithm2e}
\usepackage{hyperref}
\usepackage{float}
\restylefloat{table}

\journal{Journal of Intelligent Systems with Applications}

\begin{document}

\begin{frontmatter}

\title{SafeAccess+: An Intelligent System to make Smart Home Safer and Americans with Disability Act Compliant }

\author{Shahinur Alam}
\ead{salam@memphis.edu}
\author{Sultan Mahmud}
\ead{mmahmud@memphis.edu}
\author{Mohammed Yeasin}
\ead{myeasin@memphis.edu}
\address{Department of Electrical and Computer Engineering, The University of Memphis}
\address{206, Engineering Science Building, 
Memphis, TN 38152
}


\begin{abstract}
Smart homes are becoming ubiquitous, but they are not Americans with Disability Act (ADA) compliant. Smart homes equipped with ADA compliant appliances and services are critical for people with disabilities (i.e., visual impairments and limited mobility) to improve independence, safety, and quality of life. Despite all advancements in smart home technologies, some fundamental design and implementation issues remain. For example, people with disabilities often feel insecure to respond when someone knocks on the door or rings the doorbell. In this paper, we present an intelligent system called ``SafeAccess+" to build safer and ADA compliant premises (e.g. smart homes, banks, offices, etc.). The key functionalities of the SafeAccess+ are: 1) Monitoring the inside/outside of premises and identifying incoming people; 2) Providing users relevant information to assess incoming threats (e.g., burglary, robbery, gun violence) and ongoing crimes 3) Allowing users to grant safe access to homes for friends/family members. We have addressed a number of technical and research challenges:- developing models to detect and recognize person/activity, generating image descriptions, designing ADA compliant end-end system and feedback mechanism. In addition, we have designed a prototype smart door showcasing the proof-of-concept. The premises are expected to be equipped with cameras placed in strategic locations that facilitate monitoring the premise 24/7 to identify incoming persons and to generate image descriptions. The system generates a pre-structured message from the image description to assess incoming threats and immediately notify the users with an MMS containing the name of incoming persons or as “unknown,” scene image, and image descriptions. The completeness and generalization of models have been ensured through a rigorous quantitative evaluation. The users' satisfaction and reliability of the system has been measured using PYTHEIA scale and was rated excellent (Internal Consistency-Cronbach's $\alpha$ is 0.784, Test-retest reliability is 0.939 )
\end{abstract}

\begin{keyword}
Intelligent System, Assistive Application, Home Safety, Smart Home Technology.
\end{keyword}

\end{frontmatter}


\section{Introduction}
Building a safe smart home has been a topic of active research for decades and has received an upsurge of interest recently. People often install smart and safety equipment such as smoke detectors, poison detectors, glass-break sensors, etc., and love to turn a normal home into a smart one. However, they pay less attention to making homes compliant with the Americans with Disability Act(see ADA Compliance of SafeAccess+) and to including an intelligent system that would raise awareness to protect life/properties. According to Uniform Crime Reporting (UCR) Program conducted by FBI \footnote{https://ucr.fbi.gov/crime-in-the-u.s/2018/crime-in-the-u.s.-2018 }, an estimated 1,206,836 violent crimes and 10,208,334 property crimes occurred nationwide in 2018 and nearly \$14.5 billion worth of property was reported stolen. It is very challenging to prevent these instantaneous, violent crimes against persons/premises. However, if an intelligent system raises situational awareness, and the users can assess the incoming threats or detect ongoing suspicious activities, they can take the necessary action quickly, which will help to reduce the severity of the crime.  

	With the recent advancement in technologies, researchers and companies have developed home security solutions. However, those systems are not ADA compliant and cannot distinguish incoming threats from normal activities (described in \textbf{Problem Background} section in the supplementary document). They do not provide an easy and safe way to respond when someone knocks on the door. Moreover, those systems are often unaffordable for people with disabilities who depend on Social Security Income (SSI). People with disabilities spend most of their time at home and are vulnerable to violent crimes. To understand how safe and secure people with or without disabilities feel when they are at home alone, we have conducted a survey\footnote{https://github.com/salammemphis/Documents/blob/master/survey\%20questionnaire.docx} in Amazon mechanical Turk \footnote{https://www.mturk.com/} and followed participatory design approaches (See Participatory Design). The survey results, crime reports, safety concerns, and economic burden for the disabilities confirmed the need for developing an intelligent assistive solution. To respond to this need, we developed SafeAccess+: an end-to-end system to raise ambient/situational awareness. SafeAccess+ offers the following key benefits for private residences: 1) eliminates the burden of monitoring premises by human observers for finding suspicious activities; 2) informs the user about the person who enters the premises when the user is/is not at home; 3) assesses incoming threats; 4) generates facial descriptions of incoming persons for visually impaired; 5) allows people with disabilities to avoid the risk of navigating to the door to open/close it; 6) replaces expensive and non-intuitive doorbells; and 7) increases the independence of people with disabilities.
SafeAccess+ also provides benefits for commercial businesses as well. For example, with strategic camera placement, SafeAccess+ can provide emergency services in buildings such as banks to detect ongoing crime.

\section{Participatory Design}
The primary objective of a ``Participatory Design" is collecting the system's functional and non-functional requirements and identifying preferable modes for system interaction and notifications. It also helps to discover issues related to accessibility and usability. The participatory design emphasizes ``design for the users" and ``design with the users". To build an ADA compliant system one must collaboratively identify the user requirements and generate solutions that facilitate the identified goals. Unfortunately, requirements gathering is considered common sense and often neglected. Understanding user needs and expectation towards a product is critical to its success. One must answer the following questions before designing an assistive solution: 1) What services does the system need to offer? 2) Who are the targeted users? 3)What type of disability do they have and what is the severity of the disability? 4) How are the user's technical adaptability and cognitive ability? 5) What are the modes for the system interaction and feedback? 6) How much should the system costs? 7) How much can the user afford to pay? In order to answer those questions, we conducted a survey\footnote{https://www.mturk.com/}  with a set of 15 questions\footnote{https://github.com/salammemphis/Documents/blob/master/survey\%20questionnaire.docx}. 30 people participated in that survey (26 males, 4 female, 5 partially paralyzed, 8 visually impaired, 6 has a hearing disability and 11 has other disabilities). 12 questions were asked to collect and refine the system's functional needs and to find out user's preferences for different functional modes (feedback modes, user interaction modes with a smartphone). The collected key functional requirements from the participatory design are presented below. The other requirements are presented in the supplementary document (section 2. Participatory Design).
\\

{\textbf{Functional Requirements:}}
\begin{enumerate}
	\item Monitoring smart homes and identifying incoming persons
	\item Sending description of the incoming person and what they carry with (image description)
	\item Allowing users to remotely grant access to homes for friends and family members
	\item Providing a semi-automated option to call emergency services, friends and family members
	\item Allowing users to browse past history and video recordings from smart phone
\end{enumerate}

\section{System Design}
Design and implementation of assistive technology solutions require an understanding of user's needs as set by their disability and their ability to perform the system-aided task with minimal cognitive effort. Most of the smartphone-based assistive solutions suffer from severe criticism related to accessibility and usability \cite{dawe2006desperately}). For example, a touchscreen-based app is easy to use for sighted people, but it is not practical for the visually impaired. On the other hand, an app with a voiceover interface is convenient for the visually impaired, but people with hearing disabilities cannot use it. Therefore, we have designed the user interface with two types of interaction modes. We have followed iOS human interface guidelines \cite{iOS2019}  and Apple Accessibility Programming Guide \cite{iOSaccessibility} to make SafeAccess+ accessible for both sighted and visually impaired users. To keep the design simple and make the system effective, we have utilized the practices from Design Thinking to decide the aesthetic of the interface, System Thinking to optimize the  design and Assistive Thinking to increase accessibility. Considering the financial affordability of the users, we have developed SafeAccess+ with two modes: - 1)   standalone mode 2) integrated mode. In standalone mode, the computational unit is a low-cost raspberry pi (price \$35) and the system offers very limited features (person identification, and granting access). We have developed a CPU and Memory efficient person recognition model (see Person Recognition) for this mode. Standalone mode is designed for people with limited family income who can not buy expensive hardware. On the other hand, the integrated mode requires a large computing unit and offers full features (person identification, image description generation and granting access). The users can buy a large computing unit or rent it from a third party-provider.

The cameras need to be installed in strategic locations of a premise such as entrance/front door, back door, driveway, basement, garage, etc (see Camera Installation in the supplementary document). The users need to create a personal profile with personal information and pictures of friends/family members (see Personal Profile Creation). SafeAccess+ monitors the premises 24/7 and identifies incoming persons by matching face images with the personal profile. The system generates image descriptions and short messages from the video streams to assess incoming threats. Then, the message and scene image is sent to the user via MMS (Multimedia Messaging Service) when a person is detected. The messages are read out automatically if ``Accessibility Features" are configured in the smartphone (see Feedback Module for details). 

\begin{figure}
\includegraphics[width=\textwidth]{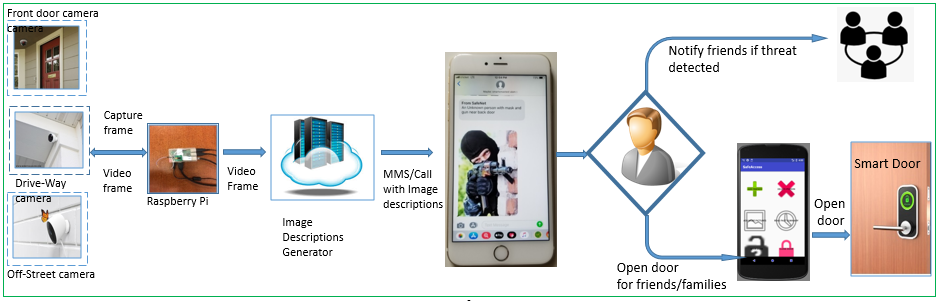}
\caption{Functional flow diagram of SafeAccess+: A raspberry pi connected to the cameras captures and encrypts the real-time streams and sends them to the Image Description Generation Unit. Image Description generator detects and identifies incoming persons from the received frame and outputs an image description. A pre-structured message generated from the image description is sent to the users.} 
\label{safeacce_arc}
\end{figure}

SafeAccess+ consists of four key components: -1) Image Description Generation module 2) Personal profile creation module 3) Smart door 4) Feedback module. The system architecture is shown in Figure \ref{safeacce_arc}. We have developed each component of the system following ADA regulations (see ADA Compliance of SafeAccess+).

\subsection{Image Description Generation Module}
\begin{figure}[htbp]
\includegraphics[width=\textwidth]{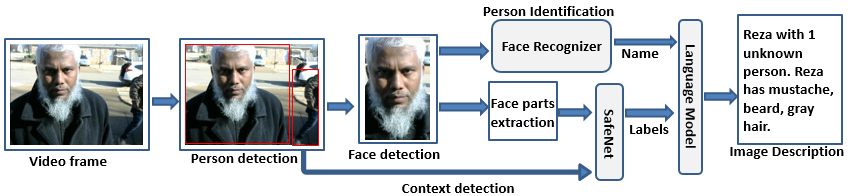}
\caption{ Workflow to generate image description} 
\label{wk_flow}
\end{figure}

We generate image descriptions to present a verbal summary of a scene to visually impaired individuals so that they can understand the scene and assess incoming threats. In addition, they will know how friends and family members look like on that visit. The content of the image description was identified from the participatory design. According to the UC Berkeley Police \footnote{https://ucpd.berkeley.edu/campus-safety/report-crime/describe-suspect } Department , survey results and outcomes from participatory design, an image description to assess incoming threats, or to identify a person may include 1) identity-name of the incoming person or ``unknown"; 2) facial description-whether a person has beard or mustache or wearing glasses/burglary-masks; 3) appearance-color of the hair, information about age, race; 4) Context- is the person carrying any harmful items such as gun, baseball bat, knife etc. The image description is generated from a cascaded process depicted in Figure \ref{wk_flow} . The first step in this process is finding the presence of a human in the monitoring zone. Then the face detection, extraction, and recognition are performed. Second, face parts are extracted from detected faces and a facial description is generated by classifying individual face parts. The colors of the hair are categorized into one of the four groups (black, brown, blond,  gray). We considered the color of hair as ``brown" or ``blond" when 60\% of the pixels from head patches have hue, saturation, and value (HSV)\footnote{https://en.wikipedia.org/wiki/HSL\_and\_HSV} between (10, 100, 20) to (20, 255, 200) and (8, 15, 50) to (20, 240, 230) respectively (experimentally found this threshold). The color ``black" is determined if the first 100 bins of intensity histogram contain 50\% of the pixels. Finally, image description is generated from class labels/words obtained from recognition outcomes and transformed into a pre-structured message using a language model. We have described each step elaborately in this article \cite{ShahinurSafeAccess+} and the following section has a brief summary. 

\textbf{Person Detection:} Finding the presence of people in the monitoring zone is critical since state-of-the-art face detection algorithms fail when the face is not visible to the camera. In those scenarios, we need to notify users someone is in the monitoring zone. We explored modern person detection algorithms and selected YOLO-V3 \cite{redmon2018yolov3} as the face detection model because of their promising performance.

\textbf{Person Recognition:} We developed two face recognition models using Support Vector Machine (SVM) trained on Local Binary Pattern (LBP) features and facial encoding extracted using a convolutional neural network. The model trained on LBP features requires low computational resources and suitable for Standalone mode (runs on raspberry pi in real-time with limited features). The model trained on both LBP features and facial encoding requires a large computing unit (GPU or multi-core machines ) to run in real-time which is suitable for Integrated mode.

\textbf{Model to Recognize Items and Facial Properties:}
We developed a convolutional neural network-based model called ``SafeNet'' to extract information about facial properties, appearances, and to recognize items that people may carry with them.

\textbf{Language Model:} We developed a rule-based language model to transform the image descriptions into pre-structured messages.
	
\subsection{Personal Profile Creation Module}
\begin{figure}[htbp]
\includegraphics[width=\textwidth]{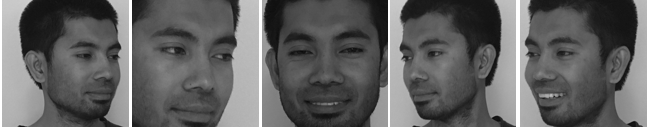}
\caption{ pictures of a person from a different view included in the profile to make the model robust against changes in view angle and position} 
\label{multiview}
\end{figure}

\begin{figure}[htbp]
\begin{center}
\includegraphics[width=\textwidth]{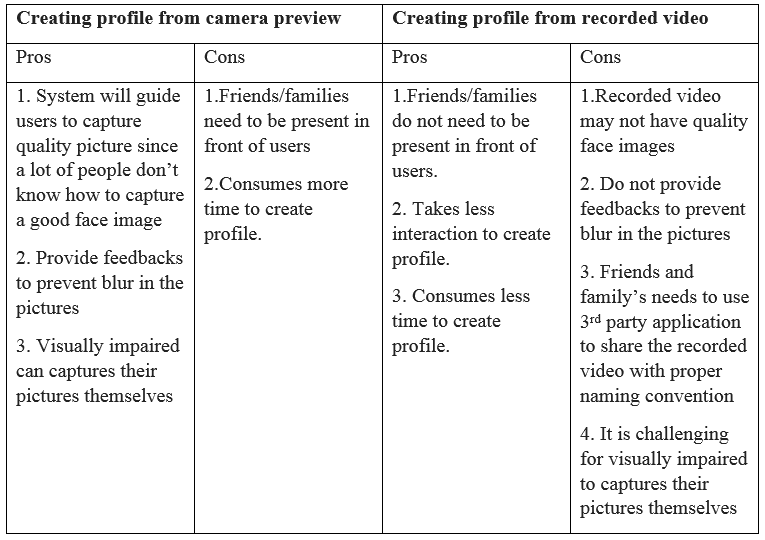}
\caption{Pros and Cons of two methods to create Personal Profile } 
\label{ppcomp}
\end{center}
\end{figure}

\begin{figure}[htbp]
 \includegraphics[width=\textwidth]{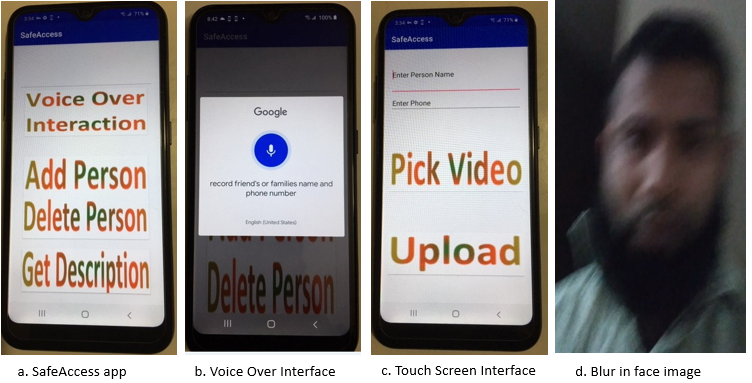}
\caption{a. SafeAccess+ App-a. main functionalities. b.Voice over interface c.Touch Screen Interface to enlist a person. d.Sample face image with blur due to abrupt movement and high speed of the camera} 
\label{app}
\end{figure}

The “Personal Profile” is a repository that contains personal information (Name, Contact) and faces images of friends/family members. The person recognition model is trained on face images stored in Personal Profile. It is advised to include face images in the personal profile with various expressions (Joy, Sadness, Surprise, Fear, Contempt, Disgust), orientation, and poses so that model can identify a person from a different view angle, position, and distances (see Figure \ref{multiview}). One of the key concerns in creating a personal profile or adding personal information to the profile is to design an ADA compliant user interface. For example, entering personal information in any online form using a personal computer/laptop may be faster. However, capturing photos are not convenient using a laptop or personal computer. Since smartphones are more accessible compared to personal computers, we developed a smartphone app (see Figure \ref{app}.a) to add a person to the profile. The app allows users to collect face images directly from the camera preview or from recorded videos shared by friends/family members. The pros and cons for both approaches are presented in Figure \ref{ppcomp}. 
 
 \textbf{Profile creation from a camera preview }: This option allows users to create a profile and capture face images directly from the camera preview. To capture face images from different views, the system guides the user to rotate smartphones around the face from left to right or right to left. The key challenges we needed to address are: 1) enabling people with disabilities, especially visually impaired individuals to take pictures of themselves and their friends/family members; 2) providing assistive guidance such as information about the position and size of a face in the camera window during image acquisitions; 3) providing information about rotational speed to prevent blur in the picture. In order to address these challenges, first, we add voiceover and touch screen interface (see Figure \ref{app}.b and \ref{app}.c) in the SafeAccess+ app so that users can choose an interaction mode based on their disabilities. 
 
\begin{figure}[htbp]
\includegraphics[width=\textwidth]{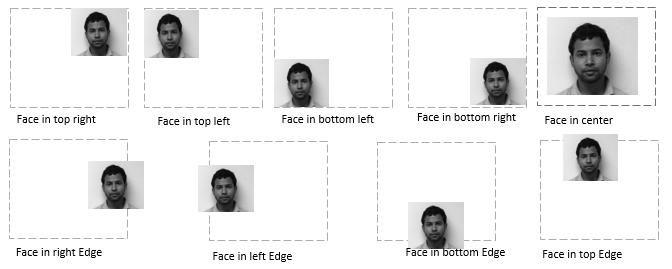}
\caption{  Guidance during image acquisition: the app reads out the position of the face in the camera window to make sure faces are not cropped. } 
\label{faceposition}
\end{figure}

Second, to make the image acquisition convenient we incorporate a face detector \cite{viola2001rapid}. It ensures the selected view has a face in the right position by providing feedback such as ``Face in top right", ``Face in top left," ``Face in bottom right," ``Face in bottom left", ``Face in left Edge", ``Face in right Edge", ``Face in top Edge", ``Face in bottom Edge" and ``Face in center" (examples shown in Figure \ref{faceposition}).  If the face is found near the edges of the camera window, it is better to change the position of the camera or person to bring the face in the center of the window. Moreover, if the person is far away from the camera and the size of the detected face is very small, the system guides the person to come closer. We calculate the positions of the face based on the four corners of the bounding box of detected faces. The underlying logic is shown in algorithm 1 (see supplementary document).
Third, the captured pictures become blurry (see Figure \ref{app}.d ) when the rotational speed is high. In order to prevent it, we provide feedback ``too fast" when the rotational speed exceeds 20 degrees per second (we found this threshold experimentally). The app automatically selects face images from the camera preview. The collected images are sent to a Webservice (see details in Utility Features: Webservices Development) which is responsible for the training/updating recognition model, versioning trained model, and data. A demo on creating Personal Profile using this approach is available at this link\footnote{https://youtu.be/ZzL5mVEuwms} (scroll to the time-point 1:22 )

\textbf{Profile creation from the recorded video}: With this option, users can enlist their friends/family members to the profile without requiring their physical presence in front of the camera. Friends/family members can record their face videos and share them using applications like ``Send Anywhere", ``SHAREit", ``AirDroid" etc.
Users hold the camera in front of the face and move the head left, right, up, down, tilted-left and tilted-right. The app interface to add a person to the profile is shown in Figure \ref{app}.c. To minimize users cognitive load and effort we have added two options to browse the recorded video: 1) The users can enter the name of friends/family members and the app will automatically find the recorded video from the phone gallery. The users need to make sure that the recorded video file is saved with the person's name as a file name so that the app's screen reader or automatic file parser can parse it; 2) Users can browse videos from the gallery. The second option may not be always convenient for people with visual impairments. The system reads the video file frame by frame and select 50 images (maximum) where the face is found in proper positions from different parts of the video. The faces are cropped from those selected images and sent to Deep Learning WebServices (See Utility Features-Webservice Development  ) to update/train the model. A demo on creating Personal Profile using this approach is available at this link\footnote{https://youtu.be/ZzL5mVEuwms} (scroll to the time-point 3:52 )
	 
\subsection{Prototype for Smart Door}
\begin{figure}
\includegraphics[width=\textwidth]{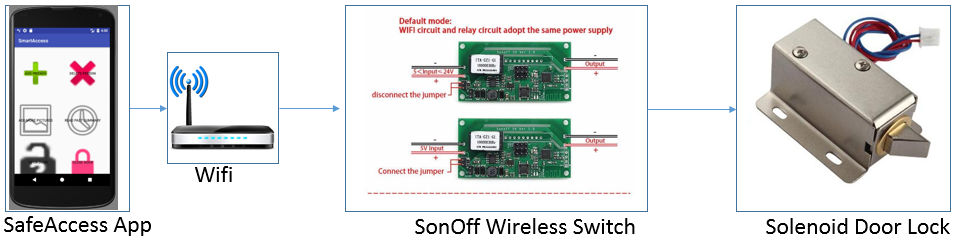}
\caption{ Prototype design of a smart door. The user sends a command from the SafeAccess+ app and the Sonoff switch turns on the solenoid lock and opens the door} 
\label{smartdoor}
\end{figure}

One of the vital components of a smart home is a safe and secure door with a smart lock. A smart lock allows the user to enter a house without requiring physical keys. Although there are many commercial smart door systems available, most of them are very expensive and people who depend on Social Security Income cannot afford them. Hence, we use a Sonoff SV Wifi-enabled switch to control solenoid lock for building a prototype and testing an end-to-end system. The prototype costs only \$12. The underlying architecture of the smart door is shown in Figure \ref{smartdoor}. Using the SafeAccess+ app users send a command to open the door and Sonoff switch turn on the lock and open the door. The door is then locked automatically after a certain time elapsed or based on user command. This time interval can be set according to users' preferences and ADA compliance\footnote{https://github.com/salammemphis\/Documents\/blob\/master\/ADAcompliance\_SafeAccess.docx}. Solenoid lock is configured to work in an inverted mode. Therefore, when there is no power in the lock, the door remains locked which allows minimizing power consumption. SonOff runs on a low voltage (5-12V DC) input power supply. To ensure the highest level of security, Sonoff can be configured to set user-defined wifi SSID (Service Set Identifier) and password. Users will be able to unlock the door from anywhere if Sonoff is connected to home internet wifi.

\subsection{Feedback Module:} Although, smartphones are equipped with accessibility features such as TalkBack, Siri, etc. many people with disabilities do not know how to use them properly. The Participatory Design discovered five modes of feedback such as TEXT/MMS, Phone Call, Alert Message, Tone, and Email. Considering the abilities of targeted users (see System Interaction and feedback Modes- in the supplementary documents), feedback is sent via MMS and/or phone call. The users can choose an option based on their technical adaptabilities and disabilities. The MMS is sent via the user's phone operator using Simple Mail Transfer Protocol (SMTP). The phone calls are made using Twilio \footnote{https://www.twilio.com/}, a 3rd party service. The user may miss the phone call or may not see the MMS timely if they are not around the phone. To address this issue they can turn ON screen-reader and assign a custom ring tone for the MMS and phone call received from SafeAccess+. The phone will play the ring tone and read out the notifications. Excessive notifications for an activity may create a cognitive load. To avoid this problem SafeAccess+ sends the first notification immediately after an activity is found and successive notifications in three minutes intervals (this threshold is found from Participatory Design). However, this interval can also be set based on user preferences. A demo on system usages and feedback messages is available at this link\footnote{https://youtu.be/ZzL5mVEuwms} (scroll to the time-point 4:56 )

\section{ADA Compliance of SafeAccess+}
	ADA compliance emphasis on Accessible Design, states that all products and services must be accessible to people with disabilities. To improve accessibility and usability of SafeAcess we followed Participatory Design (See Participatory Design ), ADA guidelines and best practices \footnote{https://developer.android.com/guide/topics/ui/accessibility} to implement each component of the system (Person Recognition, Image Description, User interface, and Feedback Mechanism). In addition, we included a set of utility features: 1) Adding voice-over and touch screen interface in our system design so that both people with hearing disability and visual impairment can use it. 2) Providing assistive guidance to enable visually impaired individuals to capture face images when they create a personal profile. 3) Making the app interface color and text compliant so that people with low vision can see it. 4) Sending the notification to users through both MMS and phone calls. Since people with vision impairments have difficulty seeing the picture attached in MMS, we include ADA compliant image descriptions for them. The guidelines followed for designing various components are described in the supplementary document (ADA Compliance of SafeAccess+-Guidelines).

\section{Utility Features}
The main purpose of the utility features is to increase efficiency and robustness of the SafeAccess+ system. The uses of various utility features and their implementation logic are presented below.

\textbf{Change Detection:}
Change detection is a process to determine whether a scene has activity or not. The captured camera frame will not have any activity unless an event occurs, such as a person enters into the monitoring area or any natural events like rain change the scene. So, if there is no activity/event in the scene, we don't need to process those frames this will reduce the burden on computational resources and network bandwidth. Moreover, it will save storage since scenes without activity will not be recorded.

SafeAccess+ uses Raspberry pi to capture the frame from the camera. Designing an efficient model/algorithm that runs on Raspberry pi to filter out frames that do not have activities is challenging. The available very deep model can not run in real-time on Raspberry pi. On the other hand, shallow models have poor person detection accuracy. Sismananda et. al. \cite{sismananda2020performance} conducted a study and showed that Yolo-Lite, an object detection model, can process 15 frame per second on raspberry pi with person detection accuracy of 61\%. Yolo-v3 \cite{redmon2018yolov3}, a comparatively deep network can detect a person with an accuracy of 99\% and process 1.15 frames per second. 

\begin{figure}[htbp]
\centerline{\includegraphics [width=\columnwidth]{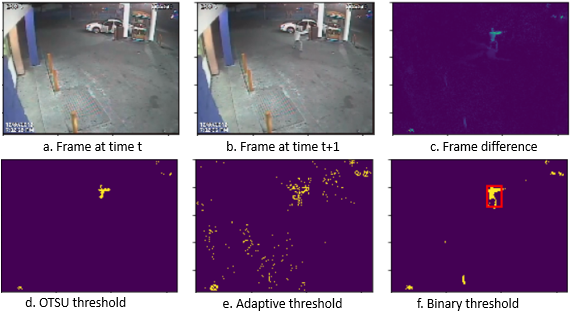}}
\caption{Change detection process: a,b) two consecutive frames c) frame difference d)outcome after applying OTSU threshold e)outcome after applying adaptive threshold f)outcome with the binary threshold. The detected person is shown in the red rectangle after applying the area threshold }
\label{change_det_process}
\end{figure}

Considering the resource constraint, computation time, and person detection accuracy, we developed a simple but efficient algorithm to find frames with activities. First, we perform change detection by subtracting consecutive frames (see Figure \ref{change_det_process}. a, b and c). The outcome  of the change detection can be  affected by  the  changes  in  contrast, brightness,  and unwanted  artifacts. The unwanted  artifacts  were  suppressed  by  applying  two  levels  of threshold on pixel values. Each pixel in the subtracted frame is compared to a threshold and if the pixel is above that threshold then the value of 255 is assigned. We examined the robustness of change detection with three types of pixel-level thresholds- a) binary or global user-defined threshold b) adaptive threshold with a Gaussian window, and c) OTSU threshold. Figure \ref{change_det_process}. d, e, and f show the outcome of the three approaches. The only difference between adaptive and binary thresholding is that the former learns the threshold from neighborhood pixels, and the latter has a global predefined threshold (threshold value 20 has been used). OTSU method\footnote{https://en.wikipedia.org/wiki/Otsu\%27s\_method} returns a threshold by minimizing intra-class intensity variance to separate background and foreground. We found from the experiment that binary thresholding provides robust change detection compared to adaptive and OTSU thresholds. Adaptive threshold is more sensitive to noise, contrast \& brightness. OTSU threshold eliminates some essential components. For example, in Figure \ref{change_det_process}.d it removed the legs of the person who is shooting someone. 

Second, we apply a binary closing and create label images by finding connected components \footnote{https://en.wikipedia.org/wiki/Connected-component\_labeling} and thresholding the area occupied by each component. Table \ref{change_detect_tb} shows performance of the change detection model based on a various threshold applied on the area. If we set a threshold greater than 32x32 to find activities/changes the model becomes very conservative and generates few false positives (high precision 0.991) and many false negatives (low recall 0.94). Provided the context we are dealing with, a high recall is more important to us compare to high precision since we don't want to miss any activity. However, when we reduced the threshold to less than 20x20, the model becomes very liberal and fails to filter out frames that do not have significant activities. Therefore, we considered 20x20 as the second level threshold. Our change detection method can process 5 frames per second.

\begin{table}[h]
\centering
\caption{Performance of Change Detection }\label{change_detect_tb}
\begin{tabular}{l|@{}c@{}|@{}r@{}|@{}c @{}}
\hline
 \textbf{Threshold on area}&\textbf{Precision} &\textbf{Recall}&\textbf{Comments}   \\ \hline
 $>32x32$ & 0.99 & 0.94 & Conservative model \\ \hline
  $<20x20$ & 0.85 & 0.99 & Liberal model \\ \hline
   20x20 & 0.96 & 0.98 & Ideal model \\ \hline
\end{tabular}
\end{table}

\begin{figure}[htbp]
\includegraphics[width=\textwidth]{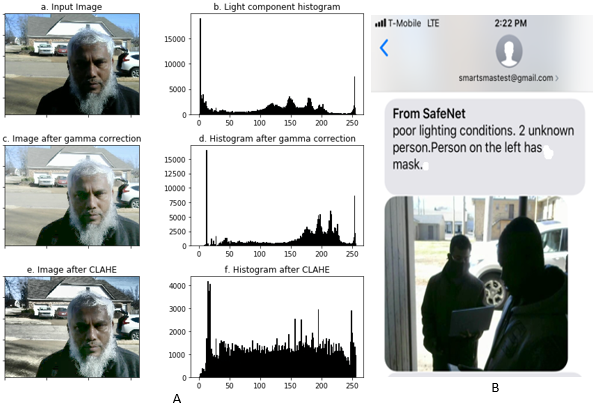}
\caption{A) Improvement of lighting condition with Gamma correction and CLAHE. B) Feedback on examining lighting condition: The lighting condition of this scene is poor because of the position of the light source, camera, and person. SafeAccess+ recognizes it and notified users } 
\label{clahe}
\end{figure}

\textbf{Lighting Condition Detection:}
The lighting condition of the surroundings is an important factor that affects any image analysis task. The person detection, recognition, and image description generation models do not work well when the lighting condition is poor. Since it is very difficult for visually impaired people to infer the lighting condition, the system automatically examines it and notifies the users (see sample scenario and notification in Figure \ref{clahe}.B). The lighting condition is examined by calculating the color histogram. If the first few bins of the histogram contain 75\% of pixels it indicates poor lighting conditions. Based on the received feedback users can turn on the external lights.

\textbf {Handling Non-Uniform Light:}
The models' performance degrades when the scene does not have uniform lighting conditions. The reason is that the training samples do not have images from various lighting conditions. For, example, Figure \ref{clahe}.a shows a sample scenario where the left part of the participant's face is dark and the right part is well-lit. In order to address this problem, we fist apply gamma correction and then apply Contrast Limited Adaptive Histogram Equalization (CLAHE) \cite{zuiderveld1994contrast}. Gamma correction is a nonlinear operation, and also known as Power Law Transformation which is used to encode and decode luminance using the following equation.

\begin{equation}
    O = I\wedge(1 / \gamma)
\end{equation}
Where I is the input image, $\gamma$ is the gamma value and O is the output image.

The $\gamma$ value less than 1 shifts the image to the darker side and $\gamma$ greater than 1 shifts the image towards the brighter side. To find the value of gamma we convert the RGB image to LAB color space and calculate histogram from the light channel, 'L'. Experimentally, we found that when the image is dark, the first 75 bins of the histogram contains more than 50\% of the pixel and we set a gamma value to 1.5. Similarly to correct over-exposed images, we set gamma value to 0.7. The results of gamma correction and CLAHE are shown for a sample image in Figure \ref{clahe}.

\textbf{Web-service Development: }
A webservice is a unit of code that exposes its functionality over the internet \footnote{https:\/\/www.tutorialspoint.com\/webservices\/what\_are\_web\_services.htm}. The primary benefit of developing a webservice is increasing interoperability. A webservice developed in one language can be used from various platforms and languages. For example, a webservice developed using Python programming can be accessed from Java, Android, Ios, etc.
To facilitate personal profile creation (see Personal Profile Creation) and model training from the smartphone app, we have developed a webservice using Python flask \footnote{https://flask.palletsprojects.com/en/1.1.x/quickstart/}. 

Collecting pictures of friends/families from the app and training model takes on average 3.8 minutes. The app should not prevent users from doing other tasks while the system transmits data or train the model. This issue was addressed by making all communication asynchronous. Google's Volley \footnote{https://developer.android.com/training/volley} framework has been utilized to make asynchronous communication convenient. 

\begin{figure}[htbp]
\includegraphics[width=\textwidth]{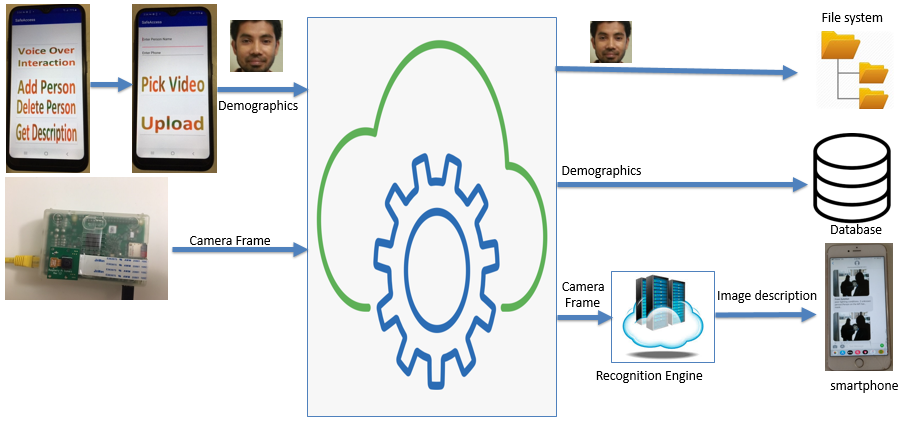}
\caption{ Webservice to facilitate personal profile creation and queries/inference} 
\label{webservice}
\end{figure}

To create a personal profile, the SafeAccess+ app sends the personal information and pictures to the webservice. Then, the webservice receives the data and saves the pictures in a file system and personal information in a database. The reason for storing the two types of information in two different structures is that the IO operation for multimedia data is faster with file system/object-store, whereas operation for other data is faster and more secure with the database system. The Raspberry Pi sends the camera streams/frames to the recognition engine via webservice call. The recognition engine processes the received frames and notifies the user. The process flow for model training and inference is shown in Figure \ref{webservice}.

\textbf{Data Security}
Data security means protecting personal information from destructive forces and from the unwanted actions of unauthorized users. Cyberattacks and data breaches are very common nowadays. If unauthorized users get access to the network or to the monitoring system they might damage properties/life. Hence, methods for transforming data should be very secure, and at the same time must not disrupt the service. Cryptography is a process for securing valuable information while transmitting it from one computer to another or storing data on a computer. Cryptography deals with the encryption and decryption of data. In order to ensure data security, we encode and encrypt the data before sending it to the ``Image Description" module from raspberry pi. It ensures that nobody can see the original information even if they hack the network. The data is encrypted using ``FerNet (Symmetric Encryption)" \footnote{https://cryptography.io/en/latest/fernet.html}. First, we generate a 32 bytes base64 password key and store it in a very secure location where only authorized users have read permission. Later, we use that password to encrypt and decrypt the data.

\begin{figure}[htbp]
\centering
\includegraphics[width=\textwidth]{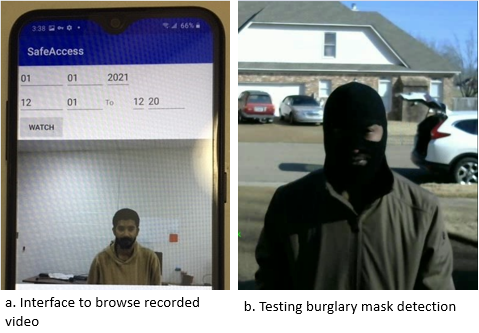}
\caption{a.Utility feature to investigate/browse recorded video to gather more information b. Sample scenario to test burglary mask detection} 
\label{investigae_video}
\end{figure}

\textbf{Browsing Recorded Videos} This feature allows users to browse and investigate past activities from the SafeAccess+ app. The users can choose a date and time-point to browse recordings. The user interface to enter date and time-points is shown in Figure \ref{investigae_video}. If there is no activity found in that selected time points ``no activity found" message will be displayed. A demo on browsing recorded videos is available at this link\footnote{https://youtu.be/ZzL5mVEuwms } (scroll to the time-point 6:26 )

\section{System Evaluation}
System evaluation is essential to make sure that the users can use the product or services and they like it, particularly if the design concept is new. The best practice for validating an assistive system is performing objective/quantitative and subjective evaluations. An objective evaluation is one that needs no professional judgment to give a score correctly, while subjective evaluation requires expert judgment to give a score \cite{suskie2004good}. We have conducted both objective and subjective evaluations of SafeAccess+ and described them in the following sections.

\textbf{Quantitative evaluation} 
The robustness of person detection, recognition, and image description generation model has been evaluated against changes in photometric and geometric properties of images, background clutters, natural artifacts, gender, and race. We  collected TEST samples from standard datasets (\citep{CelebA},\cite{Caltech},\cite{UTKFace},\cite{rgb}), movie clips, and the web to examine the computational complexity and the generalizations of the models.  
SafeAccess+ 1) detects a person with average False Acceptance and Rejection Rate of 2.39\% and 0.28\% respectively; 2) recognize a person with an average F-measure of 0.99; 3) recognize facial properties and items with an average F-measure 0.97. We have presented a detailed quantitative evaluation in this article \cite{ShahinurSafeAccess+} (see section 5. Objective Evaluation).

\textbf{Subjective Evaluation}
One of the key steps to perform a subjective evaluation of a system is to obtain a good user experience \cite{hassenzahl2006user}. User Experience is about how a person feels using a product, system, or service and emphasizes the experiential, affective, meaningful, and valuable aspects of human-computer interaction \footnote{http://en.wikipedia.org/wiki/User\_experience}. It is challenging to prepare questionnaires to collect the user's experience and score their satisfaction from all aspects. Koumpouros and colleagues \cite{koumpouros2016systematic} performed a comprehensive study to find the best approaches to measure user’s satisfaction. They recommended two scales, QUEST 2.0 and PYTHEIA for subjective evaluation. QUEST is very generic and there is not enough evidence that it has been used widely to test the assistive system. PYTHEIA is designed to test the reliability and validity of assistive solutions or services. We prepared a questionnaire with 22 statements following PYTHEIA specifications (see supplementary document, Figures 5 and 6 ). We recruited 12 participants (4 American, 4 African, 4 Asian) with a mean age of 40 years to evaluate SafeAccess+. After using SafeAccess+, the participants rate their confidence with specified statements on a 6-point Likert Scale. The Likert-scale survey is not always reliable because participants could be biased or may have low expectations toward receiving a service and their opinion may change over time. To justify the reliability of the Likert-scale survey, PYTHEIA measures: 1) internal consistency, i.e. evaluate how consistent are the results given by different questions (items) that test the latent structure of the system; 2) test-retest reliability, which is defined as the degree to which the participants maintain their opinion in the repeated measurements of the questionnaire; 3) repeatability, which is defined as the stability of participant’s responses over time, that is the ability of the system to give consistent results whenever it is used. PYTHEIA measures the internal consistency, test-retest reliability, and repeatability with Cronbach’s alpha coefficient, intra-class correlation coefficient (ICC), and Pearson's product-moment correlation coefficient respectively. We used PYTHEIA scale to measure the reliability and validity of the SafeAccess+. Moreover, we collected usages analytic such as 1) how long it takes to create a personal profile; 2) how many times the system crashes; 3) how long it takes to receive a message 

The cameras were installed at the entrance, driveways, and back doors. The participants were asked to add their family members to their personal profile. The average time it took to add a person to the profile was 3.85 minutes. Eleven participants used SafeAccess+ for an average of 1 hour. One of the visually impaired participants is using SafeAccess+ as a doorbell since April, 01, 2021. We emulated various scenarios, for example, to examine systems performance to detect burglars we wore a burglary mask and roamed in the monitoring zone (see Figure \ref{investigae_video}.b and a demo available at this link\footnote{https://youtu.be/ZzL5mVEuwms}, scroll to the time-point 5:27). In addition, we tested the system's performance to detect other harmful items such as knife, scissors, forks etc. by carrying them. In the first phase, we collected initial responses from two participants to discover any potential problems. The issues discovered and lesson learned from the first phase were: 1)the models needs to be trained with quality face images collected in well-lighting conditions; 2) the model's performance degrades when the lighting condition is not proper/uniform, especially, when images are  over-exposed or too dark. We addressed this problem using gamma correction and CLAHE \cite{zuiderveld1994contrast}(see detail in Utility Feature-Handling Non-Uniform Light Sections). In the second phase, we included ten more participants who rated the system. In order to measure the test-retest reliability and repeatability, we selected a subset of the participants who participated earlier and let them use SafeAccess+ again. The average interval of retest was 15 days.

\textbf{ Reliability and Validity TEST:}
The various reliability and validity measures are shown in Table \ref{t_reliability}. The overall Cronbach's $\alpha$ (Table \ref{t_alpha}) found from the study is 0.784, indicating sufficient consistency among the statements ($\alpha$, 0.0 =``no consistency", 1.0=``perfect consistency", greater than 0.7=``sufficient consistency"). The inter-rater agreement is excellent which indicates that the various functionalities of the SafeAccess+ received a consistent score from the participants. The Pearson correlation coefficient is 0.894, thereby indicating that the participants maintained consistent opinions over time. The test-retest measures (ICC 0.939) indicate that the system performed consistently, and users provided similar scores between two sessions. The average score received by each item/question is shown in the supplementary document( Table 1). The average score provided by each participant, Inter Item, and Intra Class (participant's score) correlations are shown in Table 2, 3, and Figure 7 respectively (see supplementary document). The average score among all questions and participants was 5.60 of 6, which shows that the participants were content with SafeAccess+.
\begin{table}[h]
\centering
\caption{Reliability and Validity Measures}\label{t_reliability}
\begin{tabular}{l|l|r}
\hline
 \textbf{Characteristics} &\textbf{Measure/Test} &
 \textbf{Value} 
 \\ \hline
 
 Internal consistency &Cronbach's $\alpha$&0.784 \\ 
  Inter Rater agreement &Absolute agreement (ICC) &0.981 \\
  Test-retest reliability  &ICC &0.939 \\
  Repeatability (rater) &Pearson's product moment r &0.894 \\
\end{tabular}
\end{table}

\begin{table}[H]
\centering
\caption{Item/Question PYTHEIA Analysis.}\label{t_alpha}
\begin{tabular}{r|r|r|r|r}
\hline
Questions&Scale Mean &Scale Variance &Corrected &Cronbach's \\
&if Item&if Item&Item-Total &Alpha if\\
&Deleted&Deleted&Correlation&item Deleted \\ \hline
Q1&112&22.889&0.551&0.766\\ \hline
Q2&111.9&23.433&0.573&0.77\\ \hline
Q3&112.4&22.933&0.422&0.771\\ \hline
Q4&112.4&25.378&-0.068&0.798\\ \hline
Q5&113.1&20.989&0.622&0.755\\ \hline
Q6&112.1&23.878&0.145&0.791\\ \hline
Q7&112.2&21.289&0.544&0.76\\ \hline
Q8&112.1&21.656&0.756&0.753\\ \hline
Q9&112.4&19.156&0.922&0.728\\ \hline
Q10&112.1&24.1&0.111&0.793\\ \hline
Q11&112.2&23.733&0.157&0.79\\ \hline
Q12&111.9&24.1&0.351&0.777\\ \hline
Q13&112&23.556&0.38&0.775\\ \hline
Q14&112.2&25.733&-0.135&0.82\\ \hline
Q15&112&21.778&0.527&0.763\\ \hline
Q16&112.5&23.167&0.406&0.772\\ \hline
Q17&112.5&23.833&0.259&0.78\\ \hline
Q18&111.9&25.433&-0.077&0.792\\ \hline
Q20&112.1&22.322&0.599&0.762\\ \hline
Q21&111.9&24.544&0.206&0.782\\ \hline
Q22&112.1&22.989&0.446&0.77\\ \hline
\end{tabular}
\end{table}

\section{Conclusion}
In this work, we have developed SafeAccess+, an end-to-end solution to build safer and ADA compliant smart homes and to increase awareness. The system monitors homes and classifies incoming persons into known vs unknown. SafeAccess+ helps people with disabilities to live independently and with dignity. In this work, we focused on gathering user's needs through a Participatory Design, developing a user-friendly assistive system and performing thorough system evaluation. The quantitative and subjective evaluations demonstrated that SafeAccess+ enhances the safety and quality of lifestyle of people with disabilities. SafeAccess+ does not help users to assess threats if an incoming person carries harmful items such as a handgun, knife and keeps them hidden. It is even very challenging for a human observer. However, robbers usually bring heavy and semi-automated guns openly to rob premises. Thus, if the monitoring cameras are installed outside and inside of a premise SafeAccess+ helps users to assess incoming threats and to detect ongoing crimes. We have improved the accessibility, usability, and robustness of SafeAccess+ so that users can have better experiences than with commercial products. However, we need to extend the scope of work in the future to meet the following requirements:
\begin{enumerate}
    \item Although, the issues related to accessibility and usability have been addressed for people with visual/hearing impairments, SafeAccess+ can not serve people who are both blind and deaf. The user interface and feedback mechanism need to be tailored so that those people can use SafeAccess+. For example, the system can provide haptic feedback in addition to voice over and text
    \item Currently, the person description has information only about facial properties. However, to describe a person completely SafeAccess+ may include more information about gender, race, age, height, etc. in the person description.
    \item SafeAccess+ distinguishes friends/family members vs unknown persons. However, when a government official such as a police officer or a postman comes to provide any service, the system should recognize them to prevent false alarm and reduce unnecessary worries of the resident.  
\end{enumerate}

\bibliographystyle{elsarticle-num}
\bibliography{Safehome}

\begin{thebibliography}{10}
\bibitem{suskie2004good}
L.~Suskie, What are good assessment practices, Assessing student learning: A
  common sense guide.
\end{thebibliography}

\end{document}


\begin{frontmatter}

\title{Supplementary Document for the Article ``Towards building Safer and Americans with Disability Act Compliant Smart Home"}

\author{Shahinur Alam}
\author{Md Sultan Mahmud}

\author[mymainaddress]{The University of Memphis}
\ead[url]{https://www.memphis.edu/}

\address[mymainaddress]{206 Engineering Science Building 
Memphis, TN 38152
}

\end{frontmatter}

\section{Language Model-Algorithm}
\begin{algorithm}[htbp]
\begin{algorithmic}
\caption{Format Message: This algorithm takes a dictionary, allmessage(v), containing list of persons with name (v.name), relative position (v.position), and words (v.desc) obtained from the recognition models associated with facial features and item they carry with. It also receives two arguments for total number of known and unknown person present in the scene. Based on the number of people present in the scene it add the details in the message.   }
\label{alg:format_message} 
\REQUIRE \textbf{Input: }  $allmessage$, $unknownprsn\_ctr$,$knownprsn\_ctr$
\STATE \textbf{Output:} $message$
\STATE $message\_prefix \leftarrow nil$
\STATE $message\_suffix \leftarrow nil$
\STATE $message \leftarrow nil$

\IF {$unknownprsn\_ctr+ knownprsn\_ctr  \geq  4 $}
\STATE  $message\_prefix, message\_suffix \leftarrow get\_person\_name\_desc(allmessage,0)$

\STATE $message \leftarrow message\_prefix +  " with " + String(unknownprsn\_ctr) + \textrm{" unknown  person."}$
\STATE  $message \leftarrow message + message\_suffix $

\ELSIF{$unknownprsn\_ctr=1 \land knownprsn\_ctr =  0 $}
\STATE  $message\_prefix, message\_suffix \leftarrow get\_person\_name\_desc(allmessage,1)$
\STATE $message  \leftarrow \textrm{"An unknown person with "} + message\_suffix$
\ELSIF{$unknownprsn\_ctr\geq 2 \land knownprsn\_ctr =  0 $}

\STATE  $message\_prefix, message\_suffix \leftarrow get\_person\_name\_desc(allmessage,2)$
\STATE $message  \leftarrow String(unknownprsn\_ctr) \textrm{" unknown person "}. + message\_suffix$" 

\ELSIF{$unknownprsn\_ctr = 0 \land knownprsn\_ctr \geq  1 $}

\STATE  $message\_prefix, message\_suffix \leftarrow get\_person\_name\_desc(allmessage,3)$
\STATE $message  \leftarrow message\_suffix$ 
\ELSE 
\STATE  $message\_prefix, message\_suffix \leftarrow get\_person\_name\_desc(allmessage,4)$

\STATE $message \leftarrow message\_prefix +  " with " + String(unknownprsn\_ctr) + \textrm{" unknown  person."}$
\STATE  $message \leftarrow message + message\_suffix $
\ENDIF

\end{algorithmic}
\end{algorithm}

\begin{algorithm}[htbp]
\begin{algorithmic}
\caption{get\_person\_name\_desc: This algorithm takes a dictionary, allmessage(v), containing list of persons with name (v.name), relative position (v.position), and words (v.desc) obtained from the recognition models associated with facial features and item they carry with. Based on the received flag it format the description to make a short message. The flag is set based on the number of people present in the scene. If the total number of people in the scene is greater than three then the name of known persons and description of the persons who have visible harmful items are included to limit the message length less than three sentence.}
\label{alg:gepersonnamedesc} 
\REQUIRE \textbf{Input: }  $allmessage$, $flag$
\STATE \textbf{Output:} $message\_prefix, message\_suffix$
 
\STATE $message\_prefix \leftarrow nil$
\STATE $message\_suffix \leftarrow nil$
\STATE $len \leftarrow length(allmessage)$

\FOR{$ \textrm{v in allmessage}$ }

    \IF{$message\_prefix != nil$}
     \STATE $message\_prefix \leftarrow message\_prefix  +" and "$
     \ENDIF
     \IF{$message\_suffix != nil$}
     \STATE $message\_suffix \leftarrow message\_suffix  +" and "$
     \ENDIF
     \STATE $message\_prefix \leftarrow message\_prefix + v.name$
     \IF{$flag =0 \land \textrm{"gun" in v.desc OR}  \textrm{" mask" in v.desc OR}   \textrm{" baseball bat" in v.desc} $}
     
      \STATE $message\_suffix \leftarrow message\_suffix  + \textrm{"Person on the "+v.position+ " has "+v.desc+"." } $
     \ENDIF
     \IF{$flag =1$}
     \STATE $message\_suffix \leftarrow message\_suffix + v.desc$
     \ELSIF{$flag =2$}
     
     \STATE $message\_suffix \leftarrow message\_suffix  + \textrm{"Person on the "+v.position+ " has "+v.desc+"." } $
     \ELSIF{$flag =3$ OR $flag =4$}
     \STATE $message\_suffix \leftarrow message\_suffix  + \textrm{ v.name+ " has "+v.desc+"." } $
     \ENDIF
\ENDFOR
\end{algorithmic}
\end{algorithm}

\bibliographystyle{elsarticle-num}
\bibliography{Safehome}


\begin{frontmatter}

\title{Supplementary Document for the Article ``SafeAccess+: An Intelligent System to make Smart Home Safer and Americans with Disability Act Compliant"}

\author{Shahinur Alam}
\ead{salam@memphis.edu}
\author{Sultan Mahmud}
\ead{mmahmud@memphis.edu}
\author{Mohammed Yeasin}
\ead{myeasin@memphis.edu}
\address{Department of Electrical and Computer Engineering, The University of Memphis}
\address{206, Engineering Science Building, 
Memphis, TN 38152
}

\end{frontmatter}

\section{Problem Background}
Although smart home technology shifted from futuristic utopias to a reality, an essential safety component that is missing in current smart homes is an intelligent integrated solution to monitor who is entering homes, assess incoming threats, and to enable people with disabilities to grant access remotely for friends and family members. The majority of the existing security solutions still depend on human observers to assess incoming threats and raise awareness. A study \cite{sulman2008effective}) has shown that human observers miss 60\% of events when they monitor nine displays. Moreover, the human observers are expensive and not practical for smart homes.

In the last decade, numerous applications have been developed to assist people with disabilities in navigation \cite {gude2013blind}, expression detection \cite{anam2014expression}, currency recognition \cite {looktel}, ambient awareness \cite{ahmed2018image}, object recognition \cite{alam2015map,kao1996object,mapelli1997role,chincha2011finding,bigham2010vizwiz}.  The technological advancement has made the object detection \cite{krizhevsky2012imagenet,girshick2014rich,girshick2015fast,sermanet2013overfeat,he2016deep} person recognition \cite {sun2015deepid3,nezami2018face} and image captioning \cite {yagcioglu2015distributed,xu2015show,ushiku2015common,vedantam2015cider,verma2014im2text,vinyals2015show}     
tasks more efficient than ever. However, researchers have paid less attention to applying those techniques to generate image descriptions from real-time video streams to assess incoming threats and raise awareness. The state-of-the-art image captioning models are not designed for these specific needs, especially, to extract information about facial properties and to recognize harmful items that a person may carry with them to commit a crime. 

In addition, the commercial product such as Vivint \cite {Vivint}, SimpliSafe  \cite {Simplisafe}, Frontpoint  \cite {FrontPointSecurity}, ADT  \cite {ADT}, Honeywell  \cite {Honeywell}, etc. depends only on motion sensors to detect activities and security breaches. For example, when someone enters the monitoring zone, it sends a push notification to the users. Then the users need to see the scene image to find out who (friends/family members/caregivers or intruders) is there. It is not suitable for people with disabilities, especially for people with vision impairments. ``Nest Hello'' \footnote{https://store.google.com/us/product/nest\_hello\_doorbell}, ``ring'' \footnote{https://ring.com/}  can recognize packages and familiar faces. However, these systems do not provide any intelligent feedbacks to assess incoming threats and are not designed to provide seamless access to homes for friends/family members.

\section{Participatory Design}
The collected non-functional requirements and users' preferences for system interaction and feedback mechanism are presented below.

\noindent {\textbf{Non-functional Requirements:}}

\begin{enumerate}
    \item \textbf{Accessibility:} The system must be accessible for the people with disabilities
    \item \textbf{Cognitive Load:} The system should not create any cognitive load
	\item \textbf{Portability:} System must be accessible from the smart phone
	\item \textbf{Maintainability:} The data should be backed up and updated periodically
	\item \textbf{Scalability:} The system must be scalable
	\item \textbf{Speed:} The system must run real-time
	\item \textbf{Data Security:} The data should be transferred securely to prevent hacking and corruption

\end{enumerate}

The following section has detailed information about the collected requirements.

\begin{figure}[htbp]
\centering
\includegraphics[width=\textwidth]{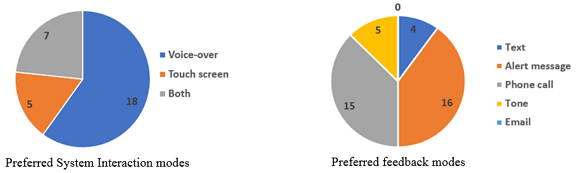}
\caption{Distribution of the preferred mode for system interaction and feedback} 
\label{pref_fd}
\end{figure}

\textbf{System Interaction and Feedback Modes:} We discovered from the crowd survey that the preferred interaction mode with the system varies based on the type of disabilities. Most of the participants with vision impairments selected the voice-over interface as the system interaction mode. On the other hand, people who have a hearing disability voted for touch screens. The distribution of the preferred system interaction mode is shown in Figure \ref{pref_fd}. The effectiveness of the system depends on how quickly we can notify users about an incoming threat. The sooner we can notify, the sooner the user can act. The distribution of the preferred feedback mode is shown in Figure \ref{pref_fd}. Most of the participants selected ``Alert Message" as a primary feedback mode since it draws more attention compare to MMS/text. A visually impaired individual suggested us to talk to vendors (Apple/Google) so that we can bypass phone's ``Do not Disturb" list for emergency alert message. The passive feedback mode such as "Email" received zero votes from the participants. Although, "Emergency Alert message" was the most popular mode for notifications it requires approval from the state or authorities. Hence, we selected MMS/Text as the primary feedback mechanism for SafeAccess+. Moreover, we trained users how to set a custom ring tone for phone calls and MMS received from SafeAccess+. It helps to draw more attention when users are not around the phone. 

 \section{Profile creation from camera preview-Algorithm}

The following algorithm is used to find the position of a face in the camera window. It generates assistive feedback so that people with visual impairment can capture face images without cropping them. The algorithm expects an image as an input and outputs the position of the face in the camera window. First, it detects faces using Viola-Jones \cite{viola2001rapid} algorithm and finds the bounding boxes of the detected faces ({x,y:top left corner}, ``width'', and ``height''). Then, it uses this information to calculate the position of a face with respect to the camera window.
	
\begin{algorithm}[H]
\begin{algorithmic}
\caption{Guidance to capture face images: The algorithm expects an image as an input and outputs the position of the face in the camera window. here, x, y (top left corner), ``width'', and ``height'' are obtained from the bounding box of detected faces. }
\label{alg:guide_prof_alg} 
\REQUIRE \textbf{Input: }  $frame$
\STATE \textbf{Output: } $face\_position$
\STATE ($x, $y, $width, height) \leftarrow get\_face\_bounding\_box(frame)$
\ENSURE $x > 0, y > 0, width > 0, height > 0$
\STATE $w \leftarrow $frame.width , $h  \leftarrow $frame.height
, $x1 \leftarrow $x - $width/2 $
, $y1 \leftarrow $y - $height/2 $
, $x2 \leftarrow $x1 + 3 * $width/2 $
, $y2 \leftarrow $y - $height/2 $
, $x3 \leftarrow $x - $width/2 $
, $y3 \leftarrow $y + 3 * $height/2 $
, $x4 \leftarrow $x + 3 * $width/2 $
, $y4 \leftarrow $y + 3 * $height/2 $

\IF {$width * height  \leq  1024 $}
\STATE $face\_position \leftarrow \textsl{``Face is small. come closer"}$

\ELSIF {$x1 \leq  0 \land  $y1 $\leq  0 $}
\STATE $face\_position \leftarrow \textsl{``Face in top left"}$
\ELSIF {$x2 \geq  w \land  $y2 $\leq  0 $}
\STATE $face\_position \leftarrow \textsl{"Face in top right"}$
\ELSIF {$x3 \leq  0 \land  $y3 $\geq  h $}
\STATE $face\_position \leftarrow \textsl{"Face in bottom left"}$
\ELSIF {$x4 \geq  w \land  $y4 $\geq  h $}
\STATE $face\_position \leftarrow \textsl{"Face in bottom right"}$
\ELSIF {$x1 \leq  0 $}
\STATE $face\_position \leftarrow \textsl{"Face in left edge"}$
\ELSIF {$y1 \leq  0 $}
\STATE $face\_position \leftarrow \textsl{"Face in top edge"}$
\ELSIF {$x2 \geq  w $}
\STATE $face\_position \leftarrow \textsl{"Face in right edge"}$
\ELSIF {$y4 \geq  h $}
\STATE $face\_position \leftarrow \textsl{"Face in bottom edge"}$
\ELSE 
\STATE $face\_position \leftarrow \textsl{"Face in center"}$
\ENDIF
\end{algorithmic}
\end{algorithm}

\section{ADA Compliance of SafeAccess+-Guidelines}
\begin{figure}[htbp]
\includegraphics[width=\textwidth]{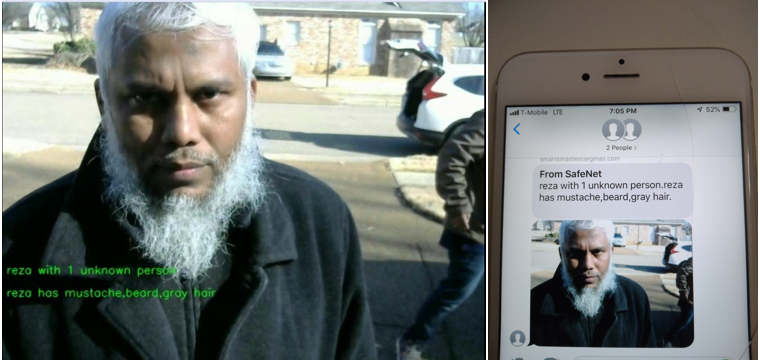}
\caption{Sample image description (left) and feedback message (right): we can see two persons in this scene. One of the known persons named "Reza" has been identified and SafeAccess+ generated an image descriptions, ``Reza with 1 unknown person. Reza has mustache, beard, gray hair"   } 
\label{reza_notification}
\end{figure}

\noindent\textbf{Guidelines for describing and recognizing a person} 
\begin{enumerate}
\item Person Identification: the system should distinguish known persons vs unknown when they enter in the monitoring zone. It will increase the comfort or awareness of the resident. SafeAccess+ identifies a person with an average F-measure of 0.98 and includes the recognition result (name or ``unknown") in the feedback message.     
\item Information about facial properties and appearance: the facial information and details of the appearance help to describe an unknown person and useful for law enforcement officials\footnote{https://ucpd.berkeley.edu/campus-safety/report-crime/describe-suspect} to investigate the crime. Moreover, visually impaired individuals do not have access to that visual information and they want to know about their friend's/family member's appearances and how they look on that visit. Hence, SafeAccess+ includes information about facial properties in the image description and  the feedback message (see Figure
\ref {reza_notification}).  
\item Information about gender, age, race and height: The information about gender, age, race, and height provides a complete description of a person. However, to limit the scope of this work those pieces of information will be included in the future version of SafeAccess+.

\end{enumerate}

\noindent\textbf{Guidelines for synthesizing Image Descriptions}
\begin{enumerate}
\item Providing a Transcript with Multimedia: The feedback message generated by SafeAccess+ contains scene images and descriptions (see Figure
\ref {reza_notification}). The visually impaired individuals can not see the scene image attached in the feedback message. Hence, SafeAccess+ includes description (transcript of multimedia) as the subject of the MMS(feedback message) so that the user can use a screen-reader and understand the scene.       
\item Image description should not be more than three lines 
\item Do not cover up graphics and other essential visual elements of the picture: To comply with this requirement the image description is added as the subject of the MMS and the scene image remains intact.
\item Use a font similar to Helvetica medium (Arial, Calibri, MS San Sherif): The feedbacks generated by SafeAccess+ are synthesized in the ``Arial" font. 
\item Have a good resolution: The quality of images should be sufficient enough so that contents are understood properly. To fulfill this requirement users are recommended to pick monitoring cameras with good resolution.  
\end{enumerate}

\noindent\textbf{Guidelines for designing the user interface}

\begin{figure}[htbp]
\includegraphics[width=\textwidth]{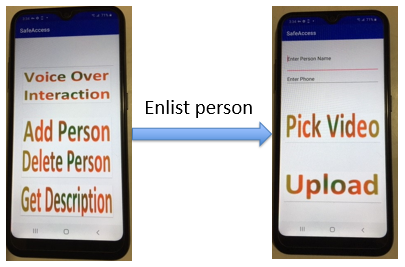}
\caption{ Color compliance of the interface: the background and item's labels were designed with a combination of red, green, and blue colors so that people who can not see a particular color can read the label} 
\label{app_touch}
\end{figure}

\begin{figure}[htbp]
\includegraphics[width=\textwidth]{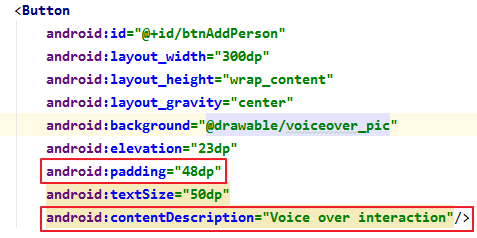}
\caption{ Content description: This Figure is to demonstrate how we added content descriptions for each element. For example, we added a content description ``Voice over interaction" for this button} 
\label{access_ui}
\end{figure}

\begin{enumerate}
    \item The User Interface should meet disability-specific needs: we have added  voice-over and touch screen interface so that both people with hearing disability and visual impairment can interact with the system (see Personal Profile Creation). The touch screen interface has been designed with a proper content description so that screen readers can verbalize the visible content and read it aloud. 
    \item Providing assistive guidance during the system use: the system provides  assistive feedback to enable visually impaired individuals to capture face images when they create a personal profile.
    
    \item The interface should be color compliant so that color-blind people can use it: to address this issue we have designed the background and label of each element with a combination of red, green, and blue color. For example, in the button label ``Add Person" (see Figure \ref{app_touch}) each letter has three colors. Hence, if a person is unable to read one color he/she will be able to read the label because of the presence of other colors.
    
    \item Designing an interface with sufficient color contrast. The interface of SafeAccess+ has been developed with sufficient color contrast so that users can distinguish the background and foreground easily. 
    
\item Adding content description (see Figure \ref{access_ui} ) to each element to make user interface accessible through the screen reader. 
    
\item Description of each element should be unique: All elements in the interface  have unique labels and content descriptions.
    
    \item The elements of the user interface should have a focusable area or touch-target size of at least 48dp x 48dp. All the elements in an interface are padded with ``48dp" (see Figure \ref{access_ui}) to make sure they do not touch the edges of the screen. Otherwise, partially blind may have difficulty seeing the contents.
\end{enumerate}

\noindent\textbf{Guidelines for Feedback mechanism}
\begin{enumerate}
    \item Feedback should not create a cognitive load. SafeAccess+ provides feedback in two scenarios: 1) during profile creation (see Personal Profile Creation); 2) and when someone enters the monitoring zone. In both cases, SafeAccess+ provides very concise feedback. 
   
\end{enumerate}

\section {Camera Installation} The key factors that need to be considered before purchasing and installing cameras are: 1) The durability of the camera considering the weather and temperature of the surroundings throughout the year. 2) Image quality and data transmission rate: the images with high resolution are better for image analysis and provide robust recognition outcomes, but transmission latency is high. We need to find a trade-off between camera resolution and data transmission latency. 3) What is the optimal view angle: how wide or narrow do we want the monitoring view be depends on the field of view angle (FOV). If the FOV of the camera is large, it can capture a wide view, but the objects in those views are very small. If the FOV is small, it can capture a small area, but objects in those views are large 4) Wireless versus wired:  wireless cameras are easy to install, but when the distance between the camera and Wi-Fi hub increases, the signal strength degrades. Hence, we need to choose a wireless versus wired camera based on the coverage area. 5) How many cameras are required: - the required number of cameras  to protect a home depends on the size of the house, coverage area, and indoor versus outdoor layout. 6) Where to place the camera: - identifying critical places to install cameras is essential because monitoring the entire area is expensive. In 2005, the law enforcement agency reported more than 2 million burglary offenses in the USA\footnote{https://www.nachi.org/burglar-resistant.htm}. The break-in points of those burglaries were the back door, front door, first-floor windows, storage area, garage, basement, and unlocked entrance. The report revealed that the first floor is more vulnerable to burglars compared to any other point of a house. Considering theft and burglary statistics, cameras can be installed at the front door, back door, off-street windows, driveways, porches, and stairways. 7) Lastly, how much does the camera cost.

	We assume that a camera will be installed at the entrance door at a height between 6-7 ft in addition to the reported break-in locations. It helps to acquire the frontal faces of the incoming person. There is a chance that people might tamper the camera if it is located at such a low height. However, there are some hidden cameras available in the market that can be used just for entrance doors. The images captured from a camera located at the top- corner of a building usually have a tilted view, and most of the time, it is difficult to align those faces/images. The state-of-the-art face alignment algorithms can frontalize faces within a limited angle.

\section{Subjective Evaluation-Questionnaire}

\begin{figure}[htbp]
\includegraphics[width=\textwidth]{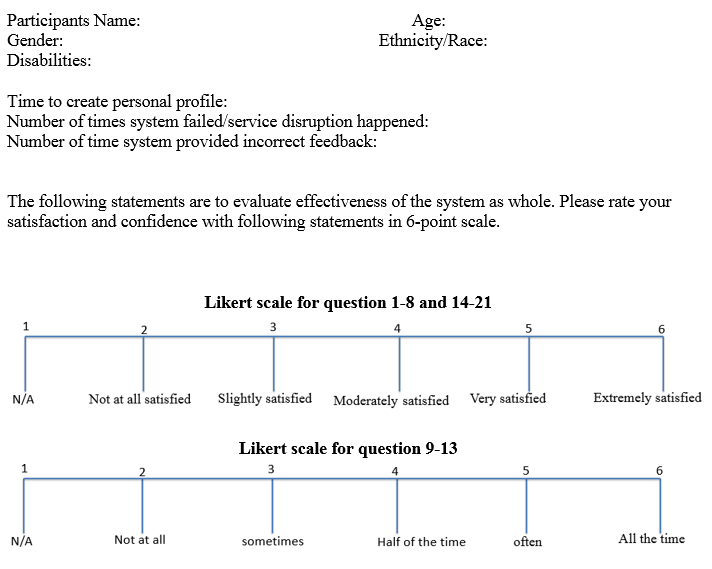}
\caption{Form: Participants information and scales to evaluate various statements} 
\label{quest1}
\end{figure}

\begin{figure}[htbp]
\includegraphics[width=\textwidth]{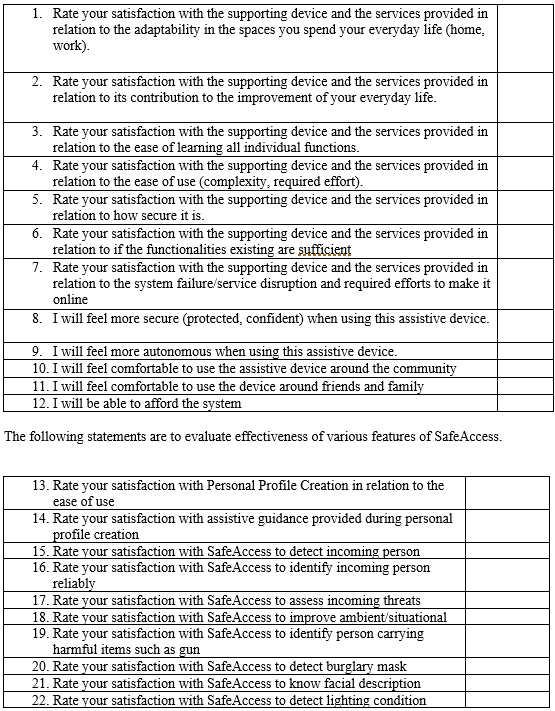}
\caption{Form: Questionnaire/Statements to evaluate SafeAccess+} 
\label{quest2}
\end{figure}

\begin{table}[H]
\centering
\caption{The average score received by each item/question from all participants}\label{t_reliability_3}
\begin{tabular}{r|r|r|r}
\hline

&Mean&Std. Deviation&N \\ \hline
Question1&5.8&0.42164&10 \\ \hline
Question2&5.9&0.31623&10 \\ \hline
Question3&5.4&0.5164&10 \\ \hline
Question4&5.4&0.5164&10 \\ \hline
Question5&4.7&0.67495&10 \\ \hline
Question6&5.7&0.67495&10 \\ \hline
Question7&5.6&0.69921&10 \\ \hline
Question8&5.7&0.48305&10 \\ \hline
Question9&5.4&0.69921&10 \\ \hline
Question10&5.7&0.67495&10 \\ \hline
Question11&5.6&0.69921&10 \\ \hline
Question12&5.9&0.31623&10 \\ \hline
Question13&5.8&0.42164&10 \\ \hline
Question14&5.6&0.84327&10 \\ \hline
Question15&5.8&0.63246&10 \\ \hline
Question16&5.3&0.48305&10 \\ \hline
Question17&5.3&0.48305&10 \\ \hline
Question18&5.9&0.31623&10 \\ \hline
Question20&5.7&0.48305&10 \\ \hline
Question21&5.9&0.31623&10 \\ \hline
Question22&5.7&0.48305&10 \\ \hline

\end{tabular}
\end{table}

\begin{table}[H]
\centering
\caption{The average score provided by participants}\label{t_reliability_2}
\begin{tabular}{r|r|r|r}
\hline

&Mean&Std. Deviation&N\\ \hline
Participant1&4.9545&1.32655&22\\ \hline
Participant2&5.5455&1.29935&22\\ \hline
Participant3&5.6364&1.29267&22\\ \hline
Participant4&5.5&1.33631&22\\ \hline
Participant5&5.1818&1.36753&22\\ \hline
Participant6&5.0909&1.306&22\\ \hline
Participant7&5.5455&1.29935&22\\ \hline
Participant8&5.2273&1.26986&22\\ \hline
Participant9&5.3636&1.329&22\\ \hline
Participant10&5.5&1.33631&22\\ \hline

\end{tabular}
\end{table}

\begin{table}[H]
\centering
\caption{Inter Item Correlation of participant's (p1, p2...p10) rating}\label{t_reliability_1}
\begin{tabular}{r|r|r|r|r|r|r|r|r|r|r}
\hline
&p1&p2&p3&p4&p5&p6&p7&p8&p9&p10\\ \hline
P1&1&0.761&0.823&0.846&0.687&0.717&0.816&0.854&0.847&0.739\\ \hline
p2&0.761&1&0.918&0.905&0.879&0.923&0.944&0.874&0.845&0.905\\ \hline
P3&0.823&0.918&1&0.937&0.847&0.867&0.946&0.923&0.857&0.882\\ \hline
P4&0.846&0.905&0.937&1&0.886&0.9&0.96&0.912&0.831&0.893\\ \hline
P5&0.687&0.879&0.847&0.886&1&0.817&0.879&0.798&0.722&0.912\\ \hline
P6&0.717&0.923&0.867&0.9&0.817&1&0.895&0.848&0.748&0.819\\ \hline
P7&0.816&0.944&0.946&0.96&0.879&0.895&1&0.931&0.845&0.932\\ \hline
P8&0.854&0.874&0.923&0.912&0.798&0.848&0.931&1&0.823&0.856\\ \hline
P9&0.847&0.845&0.857&0.831&0.722&0.748&0.845&0.823&1&0.804\\ \hline
P10&0.739&0.905&0.882&0.893&0.912&0.819&0.932&0.856&0.804&1\\ \hline

\end{tabular}
\end{table}

\begin{figure}[h]
\includegraphics[width=5.2in]{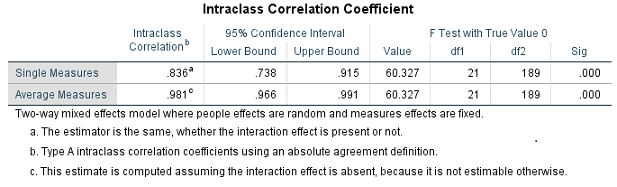}
\caption{Intra Class Correlation} 
\label{quest6}
\end{figure}

\bibliographystyle{elsarticle-num}
\bibliography{Safehome}